\documentclass[preprint]{aastex}

\shorttitle{Synchrotron Self Compton Component}
\shortauthors{Sari and Esin}

\newcommand{\g}{\gamma}
\newcommand{\gm}{\gamma_m}
\newcommand{\gc}{\gamma_c}
\newcommand{\ee}{\epsilon_e}
\newcommand{\eB}{\epsilon_B}

\eqsecnum

\begin{document}

\title{On The Synchrotron Self-Compton Emission from Relativistic Shocks \\
and Its Implications for Gamma-Ray Burst Afterglows.}
\author{Re'em Sari and Ann A. Esin\altaffilmark{1}}
\affil{130-33 Caltech, Pasadena, CA 91125; sari@tapir.caltech.edu, 
aidle@tapir.caltech.edu}
\altaffiltext{1}{Chandra Fellow}

\begin{abstract} 
We consider the effects of inverse Compton scattering of synchrotron
photons from relativistic electrons in GRB afterglows.  We compute the
spectrum of the inverse Compton emission and find that it can dominate
the total cooling rate of the afterglow for several months or even
years after the initial explosion.  We demonstrate that the presence
of strong inverse Compton cooling can be deduced from the effect it
has on the time-evolution of the cooling break in the synchrotron
spectral component, and therefore on the optical and X-ray afterglow
lightcurves.  We then show how the physical interpretation of the
observed characteristics of the synchrotron spectrum must be modified
to take into consideration this extra source of cooling, and give a
revised prescription for computing physical parameters characterizing
the expanding shock wave from the observed quantities.  We find that
for a given set of observables (synchrotron break frequencies and
fluxes) there is either no consistent physical interpretation or two
of them. Finally we discuss the prospects of directly detecting the
inverse Compton emission with Chandra.  We argue that such a detection
is possible for GRBs exploding in a reasonably dense ($n \ga 1\,{\rm
cm^{-3}}$) medium.
\end{abstract} 


\section{Introduction}
It is widely accepted that the emission from GRB afterglows is
produced in relativistic shocks, as the ejecta from an underlying
explosion expands into the surrounding medium (e.g. Piran 1999 and
references therein).  In the standard picture, the ambient electrons
are accelerated in the shock front to highly relativistic energies,
with their Lorenz factors described by a power law distribution above
some minimum value.  Besides particle acceleration, the shock is also
responsible for the creation of reasonably strong magnetic fields.
Under these conditions synchrotron radiation is produced, whose
spectrum and light curves was described by \cite*{spn98}.

Though GRBs and their afterglow are optically thin to electron
scattering for years after the initial explosion, some synchrotron
photons will Compton scatter on the shock-accelerated electrons,
producing an additional inverse Compton component at higher energies
(\citealp{pam98,wel98,tot98,chd99,pak00}; similar calculations 
in other astrophysical contexts were done by \citealp{blm77} both for
a constant density and a wind-like ambient medium). Since only a
negligible fraction of the synchrotron photons will be scattered, this
will have no direct effect on the shape of the synchrotron spectrum.
However, the ratio of inverse Compton to synchrotron luminosities is
equal to the square root of the ratio of the electron and magnetic
energy densities behind the shock front \citep*{snp96}.  Since this
number can be significantly above unity, this implies that the
electron cooling rate via inverse Compton scattering of synchrotron
photons must be considered in order to create a realistic physical
description of the GRB afterglow emission.

In their paper, \cite{pak00} give a detailed treatment of the inverse
Compton emission from GRB afterglows.  Here we extend their analysis,
concentrating on the implications of this cooling process for the
observable properties of these objects.  We start by calculating the
spectrum of the inverse Compton emission in \S\ref{spectrum}.  The
gross power-law features of this spectral component are the same as
those in \cite{pak00}.  However, here we show that the broken
power-law is not an adequate approximation above the peak frequency,
where the presence of the logarithmic terms may introduce large
corrections to the predicted flux levels.  These extra terms also
serve to smooth out the resulting spectrum and high-energy
lightcurves, eliminating sharp power-law breaks in the theoretical
predictions. 
 
In \S\ref{compton} we discuss in detail the effects of inverse Compton
cooling on the afterglow evolution.  We show that instead of the
two standard stages of evolution, corresponding to fast and slow
cooling, we now have three stages.  This occurs because the slow
cooling regime should now be divided into two, the early stage during
which inverse Compton scattering dominates the total cooling, and the
late stage when it is unimportant compared to energy loss via
synchrotron.  
In \S\ref{infer} we focus on how the physical interpretation of the
observed spectral break frequencies and the time evolution of the
afterglow emission must be modified to take into account the presence
of extra cooling via inverse Compton scattering.  In particular, we
derive a useful physical limit on a combination of the observable
parameters, measurements of which are generally the goal of afterglow
observations.  This result is then discussed in the context of
GRB~970508.

Finally, in \S\ref{direct} we discuss the possibility of  observing 
the IC emission directly.  We show that the prospects of detecting this 
spectral component are very good for GRBs which explode in a medium
with density of order $1\,{\rm cm^{-3}}$.  Our conclusions are
summarized in \S\ref{summary}.

\section{Spectrum}
\label{spectrum}
We make the standard assumption of a uniform magnetic field and a
power-law injection of electrons, with the energy distribution given
by $N(\g) \propto \g^{-p}$, behind the expanding shock front.  Then
the general shape of the afterglow spectrum is determined by the
relationship between the electron cooling time and the system
dynamical time.  Below we consider two limiting cases: when the
majority of the injected electrons can cool on the dynamical time of
the system (fast cooling) or when the cooling affects only the
electrons in the high energy tail of the distribution (slow
cooling). 

\subsection{Slow Cooling}
\label{scspec}

In this case the electron energy distribution around the minimum
injection energy, $\gm$, is not affected by the cooling. However,
electrons with Lorenz factors above some critical value, $\gc$,
radiate significant fraction of their energy on the dynamical time.
Since the ratio of the cooling time of the electron with Lorenz factor
$\g_e$ to the dynamical time is $(\gc/\g_e)$, the electron
distribution above $\gc$ become steeper by one power.  Therefore, the
resulting electron energy distribution is :
\begin{equation}
\label{ngammasc}
N(\g)\propto \left\{ \begin{array} {ll} 
\g ^{-p}, & \mathrm{if} \quad \gm<\g <\gc \\ 
\g ^{-p-1}, & \mathrm{if} \quad \gc<\g. 
\end{array} \right. 
\end{equation}

As described by \cite{spn98}, the synchrotron emission from such
distribution of electrons can be approximated by a broken power-law
spectrum with three characteristic break frequencies.  One is the
self-absorption frequency, $\nu_a$, below which the system becomes
optically thick.  The other two are the peak frequencies of the
emission from the electrons with the characteristic Lorenz factors
$\gm$ and $\gc$, denoted $\nu_m$ and $\nu_c$, respectively.  Here we
assume that $\nu_m$ and $\nu_c$ are greater than $\nu_a$.  At
frequencies $\nu < \nu_a$, self-absorption is important and the specific
flux is proportional to $\nu^{2}$, while between $\nu_a$ and $\nu_m$
the synchrotron emission grows more slowly with $f \propto \nu^{1/3}$.
The specific flux peaks at $\nu_m$ and then decreases as $\nu
^{-(p-1)/2}$ for frequencies $\nu_m < \nu <\nu_c$; above $\nu_c$ it
decreases as $\nu^{-p/2}$.

Using this synchrotron spectrum as the source of seed photons, we
computed the resulting inverse Compton emission, by integrating over
the differential scattering cross section and over the electron energy
distribution given by Eq. \ref{ngammasc} (this calculation is
discussed in detail in appendix~A).  We find that, like the original
synchrotron spectrum, the upscattered component consists of four
segments, with the breaks at three characteristic frequencies:
$\nu_a^{IC} \simeq 2 \gm^{2}\nu _{a}$, $\nu_m^{IC} \simeq 2 \gamma
_{m}^{2} \nu_m$, and $\nu_c^{IC} \simeq 2 \gc^{2} \nu_c$.  At
frequencies $\nu < \nu_m^{IC}$, the main contribution to the IC
spectral component comes from synchrotron photons scattered by the
electrons with Lorenz factor $\gm$.  In this region, the spectrum is
well approximated by two power-law segments.  The specific flux
initially increases linearly with frequency up to $\nu_a^{IC}$; this
spectral slope is determined by the shape of the scattering cross
section (see appendix A).  It then continues to rise as $\nu ^{1/3}$
up to its peak at $\nu = \nu_m^{IC}$.

In the frequency range $\nu_m^{IC} < \nu < \nu_c^{IC}$, electrons with
a range of Lorenz factors between $\gm$ and $\gc$ contribute equally
to the emission at each frequency.  Because of this, the flux drops
off roughly as $\nu^{-(p-1)/2}$, but with two major differences from
the original synchrotron spectrum.  Firstly, the region with this
spectral slope extends over double the frequency range (in logarithmic
units) of the corresponding synchrotron component,
i.e. $\nu_m^{IC}/\nu_c^{IC}=(\nu_m/\nu_c)^2$. Secondly, contributions
from different electron energies generate a logarithmic term, which
peaks at $\nu=\sqrt{\nu_m^{IC}\nu_c^{IC}}$, increasing the flux there
by a factor of order $\ln{(\nu_c/\nu_m)}$ above that of the the underlying
power-law (see Eq. [\ref{fc2}]).

Similarly, at frequencies above $\nu_c^{IC}$, the emission is composed
of equal contributions from electrons with a range of Lorenz factors $\g_e >
\gc$.  Here specific flux drops off as $\nu ^{-p/2}$, but also with an
addition of a logarithmically growing term of order $\ln{(\nu/\nu_c^{IC})}$ 
(see Eq. [\ref{fc2}]).

The inverse Compton spectral component, approximated as a broken
power-law and normalized using Eq. (\ref{fratio}), below is shown in
Fig. \ref{fig1}a together with the more ``exact'' spectrum obtained by
computing the integrals in Eq. (\ref{fc}).  To emphasize the
importance of IC emission as a cooling mechanism, we plotted $\nu
f_{\nu}$ vs. $\nu$, which gives a measure of energy emitted per
logarithmic frequency interval.  The comparison of the two curves
shows that the power-law approximation does a very good job at
frequencies below $\nu_m^{IC}$, but becomes somewhat inadequate at
higher frequencies.  In fact, our calculation shows that at
frequencies $\nu > \nu_m^{IC}$, the spectrum of the IC emission has 
a continuously varying slope, without identifiable spectral
breaks.  Moreover, one must keep in mind the the observed IC spectral
component is expected to be even smoother, since in our calculations
we assumed a broken power-law distribution of the synchrotron seed
photons and used a simplified description of the scattering cross
section (see appendix A).

Another feature of the detailed spectrum, which cannot be deduced from
the power-law approximation, is that for small values of $p$ the IC
energy output actually peaks well above $\nu_c^{IC}$ (which is the
peak emission frequency for the approximate power-law spectrum).  It
is easy to show that this feature is due to the presence of the
logarithmic correction term. It persists for $2 < p \la \sqrt{6}$,
which includes the range of values deduced from observations
\citep*[e.g.][]{wig99,sph99,frw00}.

\subsection{Fast Cooling}
\label{fcspec}
In this regime all of the injected electrons are able to cool on the
dynamical time scale of the afterglow.  Therefore, there is a
population of electrons with Lorenz factors below the injection
minimum $\gm$, and the resulting electron distribution takes the form
\begin{equation}
\label{ngammafc}
N(\g)\propto \left\{ \begin{array} {ll} 
\g ^{-2}, & \mathrm{if} \quad \gc<\g <\gm; \\ 
\g ^{-p-1}, & \mathrm{if} \quad \gm<\g. 
\end{array} \right. 
\end{equation}

The synchrotron spectrum has the same three characteristic break
frequencies, though now $\nu_c < \nu_m$, and consists of four
power-law segments.  Below $\nu_a$ the emission is optically thick and
the specific flux increases as $\nu^{2}$, and in the range $\nu_a < \nu
<\nu_c$ we have $f_{\nu} \propto \nu ^{1/3}$.  The flux peaks at $\nu_c$ and
declines as $f \propto \nu ^{-1/2}$ for $\nu_c < \nu < \nu_m$ and as
$f\propto \nu^{-p/2}$ above $\nu_m$. 

A spectrum of the inverse Compton emission in this regime is plotted
in Fig. \ref{fig1}b. Its general features are very similar to those of the slow
cooling case described above. The specific flux is linear in frequency
below $\nu_a^{IC}$ and continues as $f \propto \nu^{1/3}$ up to its
peak at $\nu_c^{IC}$, in the range $\nu_c^{IC} < \nu < \nu_m^{IC}$ it
declines as $f \propto \nu ^{-1/2}$ and drops off with $f \propto \nu
^{-p/2}$ above $\nu_m^{IC}$.  Since most of the contribution to the IC
emission below $\nu_c^{IC}$ comes from the lowest energy electrons, a
broken power-law is a fairly good approximation to the spectrum in
that region.  At higher frequencies electrons with a range of Lorenz
factors contribute to the emission, creating additional logarithmic
terms (see Eq. [\ref{fc2fc}]), which smooth out the breaks in the
spectral slope.

\subsection{Peak Flux and Luminosity Ratios}
\label{ratios}

The strength of the inverse Compton emission relative to synchrotron
can be estimated by considering the ratio of specific fluxes measured at
the peak of the respective spectral components (
peak of $f_{\nu}$ rather than $\nu f_{\nu}$).  
We denote these peak fluxes as $f_{max}$ and
$f_{max}^{IC}$ for the synchrotron and inverse Compton components,
respectively.  As described in \S\ref{scspec} and \S\ref{fcspec}
above, the former peaks at $\nu_m$ and the latter at $\nu_m^{IC}$ in
the slow cooling regime; and at $\nu_c$ and $\nu_c^{IC}$,
respectively, in the fast cooling regime. In both cases the number of
electrons that contribute to the radiation around the maximum is
proportional to the total number of electrons in the system,
$N$. Therefore the ratio of the two fluxes is simply
\begin{equation}
\label{fratio}
\frac{f_{max}^{IC}}{f_{max}} \sim \frac{\sigma_T N}{4 \pi R^{2}} \sim
\frac 1 3 \sigma_T n R =2 \times 10^{-7} n_1 R_{18},
\end{equation}
where the last two terms assume that the electrons have been collected
from an ambient medium with particle density $n$ over a distance $R$.
For a range of values of $n$ and $R$, which are relevant to the
observable GRB afterglows, the flux ratio above is always much less
than unity.

To assess the relative importance of contributions from the inverse
Compton and synchrotron emission to the total cooling rate of the
electrons, we can compute the ratio of the total energies emitted via
these two mechanisms.  Instead of integrating over the entire
spectrum, we evaluate $\nu f_{\nu}$ at the peaks of the two spectral
components and simply compute a ratio of the two values.  Note that
taking $L_{syn} \sim \nu_c f_{\nu} (\nu_c)$ and $L_{IC} \sim
\nu_c^{IC} f_{\nu} (\nu_c^{IC})$ is not in general a good assumption, since
direct spectral integration gives luminosities a factor $>$10 greater
than this estimate.  However, since the shapes of the synchrotron and
IC spectral components are similar, the correction factors nearly
cancel out.  

In the slow cooling regime, the energy emitted via synchrotron peaks
at $\nu_c$.  The inverse Compton emission can reach its maximum energy
output at $\nu \gg \nu_c^{IC}$ for small values of $p$ (see
\S\ref{scspec}), however, since $\nu f_{\nu}$ is nearly flat in this
region of the spectrum, for the purposes of this estimate we use
$\nu_c^{IC} f_{\nu} (\nu_c^{IC})$.  Then the luminosity ratio during
the slow cooling regime is
\begin{equation}
\label{lratiosc}
\frac{L_{IC}}{L_{syn}}\sim \frac 2 3 \sigma_T n R \gc^{2} 
\left(\frac{\gc}{\gm}\right)^{1-p} 
\end{equation}

In the fast cooling regime the synchrotron and the IC energy emission peak at
$\nu_m$ and $\nu_m^{IC}$, respectively, and the luminosity ratio is given by
\begin{equation}
\label{lratiofc}
\frac{L_{IC}}{L_{syn}} \sim \frac 2 3 \sigma_T n R \gc \gm.
\end{equation}

\section{Afterglow Evolution with Strong Compton Cooling}
\label{compton}
Throughout most of the early afterglow evolution, inverse Compton
emission typically dominates the total cooling rate, and therefore it
determines the value of break energy in the electron energy
distribution.  This has a significant effect on the time evolution of
both the synchrotron and the inverse Compton component.

\subsection{Luminosity ratio}
\label{lumratio2}
The ratio of the inverse Compton to synchrotron luminosity can be
computed in a more general way \citep*{snp96}, that does not deal with
the details of the spectrum, but depends only on the underlying physical
properties of the expanding shock wave .  We generalize the derivation
given by \cite{snp96} to describe both fast and slow cooling regimes by
introducing a parameter $\eta$, equal to the fraction of the electron
energy that was radiated away (via both synchrotron and IC emission).
Then the ratio of luminosities, in the limit of single scattering, is
given by
\begin{equation}
\label{x}
x\equiv \frac{L_{IC}}{L_{syn}} = \frac{U_{rad}}{U_{B}} = 
\frac{U_{syn}}{U_{B}} = \frac{\eta U_{e}/(1+x)}{U_{B}} =
\frac{\eta \ee}{\eB (1+x)}, 
\label{ICtoSYN}
\end{equation}
where $U_{syn}$, $U_B$ and $U_e$ are the energy density of synchrotron
radiation, magnetic field and relativistic electrons, respectively.
Note that in general $U_{syn} = \eta \beta U_e/(1+x)$, where $\beta$
is the velocity of material behind the shock front (in the frame of
the shock); however, for a relativistic shock $\beta$ is of order
unity. The parameters $\ee$ and $\eB$ are defined as fractions of the
total explosion energy that go into accelerating electrons and amplifying
magnetic fields behind the shock front, respectively.

Solving Eq. (\ref{x}) for $x$ we obtain
\begin{equation}
x=\frac {-1+\sqrt{1+4 \frac {\eta\epsilon_e} {\epsilon_B} }} {2}.
\label{exactx}
\end{equation}
This solution has two interesting limits:
\begin{equation}
\label{x2}
x = \left\{ \begin{array} {ll} 
\frac{\eta \ee}{\eB}, & \mathrm{if \quad}
\frac{\eta \ee}{\eB} \ll 1, \\ 
\left(\frac{\eta \ee}{\eB}\right)^{1/2}, & \mathrm{if \quad}
\frac{\eta \ee}{\eB} \gg 1. 
\end{array} \right.
\label{approxx} 
\end{equation}
Hereafter, unless specified otherwise, we always use these approximate
solutions for $x$.

Clearly, if $\eta \ee/\eB \ll 1$, the inverse Compton cooling rate is
unimportant compared to synchrotron, and if $\eta \ee/\eB \gg 1$, it
dominates the total emission.  Using these expressions, the relative
importance of IC emission can be evaluated using only the fundamental
properties of the expanding shock, and the ability of the electrons to
cool.  We can reduce this result to that in
Eqs. (\ref{lratiosc}, \ref{lratiofc}) above by noting that (i) for
slow cooling $\eta=(\gc/\gm)^{2-p}$ and for fast cooling $\eta=1$;
(ii) $\gm \sim \ee U /m_e c^2 n$, where $U$ is the total energy
density behind the shock; (iii) $\gc \sim m_e c^2 /\sigma_T c \eB U t$
if $\eta \ee/\eB \ll 1$, and $\gc \sim m_e c^2 /x \sigma_T c \eB U t$
if $\eta \ee/\eB \gg 1$.

\subsection{Effects on Synchrotron Spectrum and Lightcurves} 

Inverse Compton scattering can, in principle, affect the observed
spectrum in three different ways (see also discussion in Sari, Narayan
and Piran 1996).  In the first place, it reduces the number of seed
photons, changing the overall normalization of the synchrotron
spectral component.  This is important only if the Thompson optical
depth is larger than unity, and is therefore negligible in GRB
afterglows.  Secondly, IC scattering produces an additional emission
component at higher frequencies, which we described it
\S\ref{spectrum}.  Finally, when the inverse Compton emission
dominates the overall cooling of the electrons, it reduces the energy
available for synchrotron radiation.  Consequently, the cooling break
energy of the electron distribution, $\gc$, is reduced from its
standard (synchrotron only) value by a factor $(1+x)$, i.e. in all 
our calculations we must replace $\gc \rightarrow \gc/(1+x)$.  It is the
results of this latter effect that we discuss below.

From Eq. \ref{x2} it is clear that if $\ee<\eB$, the IC power is
never larger than the synchrotron power, since by definition $\eta \le
1$. However, current estimates based on afterglow observations, show
that $\eB<\ee$; in fact in several objects the ratio $\ee/\eB$ seems
to be close to a hundred \citep{wig99,gps99}.  This implies that the
magnetic field is relatively weak, and IC emission is important as
long as $\eta \ga 0.01$.

Assuming that $\ee > \eB$, the evolution of a GRB afterglow must now
be divided into three regimes.  In the first (fast cooling) stage,
most of the electrons energy is lost, so that $\eta=1$ and IC emission
dominates over synchrotron by a time-constant factor $\sqrt{\ee/\eB}$.
Later, during the slow cooling period, $\eta$ decreases and so is the
importance of IC scattering, though it still dominates the total
cooling.  Finally, $\eta$ becomes so small that only synchrotron
emission is important; this is the traditional slow cooling regime.
Below we discuss each of these stages in detail.  Note that all our
numerical results are computed in terms of $\epsilon_{B,-2} =
\eB/0.01$ and $\epsilon_{e,.5}= \ee/0.5$.

Since temporal evolution of the afterglow emission depends on the
external density profile, the results are different in the case of the
uniform ambient medium and the wind scenario, in which the density
decreases as $\rho \propto r^{-2}$. In this section we will present
the results for the constant ambient density, and discuss the
wind-like density profile in appendix B.
 
\subsubsection{Fast cooling} 
\label{fcIC}
During the fast cooling stage, IC scattering increases the radiation
losses of each electron by a factor $1+x \simeq x \simeq
\sqrt{\ee/\eB}$.  Therefore, the cooling frequency $\nu_c \propto
\gc^2$ is reduced from its synchrotron-only value by a factor $x^2
\simeq \ee/\eB$. Moreover, since low energy synchrotron emission is
dominated by the electrons near the cooling break, the synchrotron
self-absorption frequency increases by a factor $x \simeq
\sqrt{\ee/\eB}$.  Because both $\ee$ and $\eB$ remain constant in time,
these changes do not affect the time evolution of the synchrotron
spectrum.

Since IC scattering increases the total cooling rate, it prolongs the
duration of the fast cooling regime. For a constant ambient density the
transition to slow cooling is delayed by a factor $\ee/\eB$, if $\ee >
\eB$.  The correct transition time then becomes
\begin{equation}
t_0^{IC}=170\, (1+z) \ee^3 \eB E_{52} n_1\,{\rm days} = 
6.3\, (1+z) \epsilon_{e,0.5}^3 \epsilon_{B,-2} E_{52} n_{1}\, {\rm hours},
\end{equation}
where $E = E_{52} \times 10^{52}\,{\rm erg\, s^{-1}}$ is the total
energy of the explosion, and $n = n_1\times 1\,{\rm cm^{-3}}$ is the
ambient electron density.  By comparison, if $\ee/\eB<1$, IC
scattering is never an efficient cooling mechanism and we get the
usual (synchrotron only) expression:
\begin{equation}
t_0=170\, (1+z) \ee^2 \eB^2 E_{52} n_{1}\, {\rm days}.
\end{equation}

\subsubsection{Slow IC-Dominated Cooling} 
\label{scIC}
During the slow cooling stage only the cooling frequency is affected by
IC cooling.  The parameter $\eta$ is no longer equal to unity, so
during this intermediate regime $\nu_c$ is reduced by a factor
$(1+x)^2 \simeq x^2 \simeq \eta \ee/\eB$, as long as $x > 1$.  It is
important to note that though $\eta$ decreases with time, its
instantaneous value can be estimated directly from the observed
synchrotron spectrum alone, since in the slow cooling regime
$\eta=(\gc/\gm)^{2-p}=(\nu_c/\nu_m)^{-\frac{p-2}2}$.  Since the value
of $p$ is estimated to be in the range of $2.2-2.4$ the decrease of
$\eta$ is very slow and inverse Compton scattering remains a dominant
cooling process for a long time.

Without inverse Compton cooling, $\nu_c\propto t^{-1/2}$ and $\nu_m \propto
t^{-3/2}$ so that their ratio is $\nu_c/\nu_m = t/t_0$. Taking IC emission
into account we have $\nu_c/\nu_m \simeq (t/t_0) x^{-2}$, and $x$
evolves with time according to the following equation,
\begin{equation}
x=\sqrt{\frac {\eta \ee}{\eB}}=
\left(\frac{\nu_c}{\nu_m}\right)^{-\frac{p-2}4}\sqrt{\frac {\ee}{\eB}}\simeq
\left(\frac{t/t_0}{x^2}\right)^{-\frac{p-2}4}\sqrt{\frac {\ee}{\eB}},
\end{equation}
which gives
\begin{equation}
\label{xtime}
x\simeq\left(\frac{\ee}{\eB}\right)^{\frac{1}{(4-p)}}
\left(\frac{t}{t_0}\right)^{-\frac { (p-2)} {2(4-p)}} =
\sqrt{\frac{\ee}{\eB}}
\left(\frac{t}{t_0^{IC}}\right)^{-\frac{(p-2)}{2(4-p)}}.
\end{equation}
One should keep in mind however, that in deriving the equation above
we used the relativistic time scalings for the frequency ratio
$\nu_c/\nu_m$. Once the fireball is no longer relativistic, our basic
expression for $x$ (Eq. \ref{x}) is no longer valid. In the non-relativistic
case $x$ decreases as $t^{-3/10}$ for $p=2$, and even more rapidly for
larger $p$, considerably faster than implied in Eq. (\ref{xtime}).
Therefore, IC cooling ceases to be dominant shortly after the
relativistic stage is over.

\subsubsection{Slow Synchrotron-Dominated Cooling} 
\label{scnoIC}
The third stage of afterglow evolution, when $x < 1$ and the IC
cooling rate is weaker than the synchrotron cooling rate and
therefore, has no effect on the synchrotron spectrum, begins at time
\begin{equation}
t^{IC}=t_0^{IC} \left( \frac {\ee} {\eB} \right)^ {\frac {4-p}
{(p-2)}}=3\, {\rm years} \left( \frac {\epsilon_{e,.5}} {\epsilon_{B,-2}} 
\right)^{\frac {4-p} {p-2}} \epsilon_{e,.5}^3 \epsilon_{B,-2} E_{52} n_{1},
\label{tIC}
\end{equation}
with the numerical coefficient computed for $p=2.3$.  Note that this
expression is very sensitive to the value of $p$, and goes to infinity
when $p \rightarrow 2$.   

Equation \ref{tIC} implies that for $\ee/\eB \ga 10$, the IC
scattering dominates total cooling over the whole relativistic stage
of afterglow evolution, and therefore, over the time period containing most
current observations.

\subsection{Inverse Compton Lightcurves}
Equations describing light curves of inverse Compton emission in the
regimes of fast cooling and slow cooling with negligible IC (as
defined in \S\ref{fcIC} and \S\ref{scnoIC}) were given in
\cite{pak00}, so we omit them here. One must keep in mind however,
that these equations were derived using the broken power-law
approximation to the IC spectral component, and so must be used with
caution.  During the intermediate stage (see \S\ref{scIC}), when
$\nu_c > \nu_m$ but the total cooling rate is dominated by IC
scattering, the break frequencies of the IC spectral component and its
peak flux evolve with time in the following way
\begin{eqnarray}
\label{nutime}
\nu_a^{IC} &=& 2 \gm^2 \nu_a \propto t^{-3/4}, \\ \nonumber
\nu_m^{IC} &=& 2 \gm^2 \nu_m \propto t^{-9/4}, \\ \nonumber
\nu_c^{IC} &=& 2 \gc^2 \nu_c \propto (1+x)^{-4}t^{-1/4} 
\propto t^{-1/4+\frac{2(p-2)}{4-p}}, \\ \nonumber
f_m^{IC} &\sim& \sigma_T R n f_m \propto t^{1/4}.
\end{eqnarray}
For a range of values of $p$ consistent with constraints from
observations, the break frequency $\nu_c^{IC}$ remains nearly constant
or increases with time: $\nu_c^{IC} \propto t^{-0.03}$ for $p=2.2$ and
$\nu_c^{IC} \propto t^{0.25}$ for $p=2.4$.  It is therefore unlikely
that it can ever be observed directly with X-ray instruments (see
Fig. \ref{fig1}).

Using the expressions above and the detailed spectrum described in the
appendix, one can obtain the lightcurve of the IC emission at any
frequency. As we have emphasized earlier, taking only the broken
power-law approximation to the IC spectrum can miss considerably, and
can be used only to get a very rough estimate, given below
\begin{equation}
f_{\nu}^{IC} \propto \left\{ \begin{array} {ll} 
t^{9/4}, & \mathrm{if \quad} \nu < \nu_{a}^{IC},\\
t^{1}, & \mathrm{if \quad} \nu_a^{IC} < \nu < \nu_{m}^{IC},\\
t^{-\frac {9p-11} 8}, & \mathrm{if \quad} \nu_m^{IC} < \nu < \nu_{c}^{IC},\\
t^{-\frac {9p-10} 8+\frac{p-2}{4-p}}, & \mathrm{if \quad} \nu_c^{IC} < \nu.
\end{array} \right.
\label{IClightcurve}
\end{equation}

The equation above differs from that given by Panaitescu and Kumar
above the frequency $\nu_c^{IC}$. Their expression is valid only it
late times, when IC cooling is not important, while
Eq.~(\ref{IClightcurve}) is valid during the intermediate stage of
afterglow evolution (slow cooling dominated by IC). As we stressed
earlier, for typical GRB parameters, this intermediate stage lasts
throughout the relativistic evolution of the fireball.

\section{Inferring Afterglow Parameters from the Synchrotron Spectral 
Component}
\label{infer}
Even if the IC emission is not directly observed, its presence affects
how we interpret the properties of the observable synchrotron spectral
component. The general shape of the synchrotron emission is not
affected by the IC scattering.  However, when this extra source of
cooling is taken into account, it lowers the predicted value of the
cooling frequency, $\nu_c$ by a factor $(1+x)^2$. In addition to this,
since during the fast cooling stage the self absorption frequency,
$\nu_a$, depends on the the cooling frequency, in this regime its
value increases by a factor $(1+x)$. The resulting modified
expressions for the break frequencies and the peak flux of the
synchrotron spectrum are listed below:
\begin{eqnarray}
\label{nua}
\nu _{a}^{slow}&=& 3.6\,{\rm GHz\,} (1+z)^{-1} \epsilon_{e,0.5}^{-1} 
\epsilon_{B,-2}^{1/5} E_{52}^{1/5} n_1^{3/5},  \\
\nu _{a}^{fast} &=& 0.15\,{\rm GHz\,} (1+z)^{-1/2} \epsilon_{B,-2}^{6/5}
E_{52}^{7/10} n_1^{11/10} t_{day}^{-1/2} (1+x), \\
\nu_m&=&5 \times 10^{12}\, {\rm Hz\,} (1+z)^{1/2} \epsilon_{B,-2}^{1/2}
\epsilon_{e,0.5}^2 E_{52}^{1/2} t_{day}^{-3/2}, \\
\nu_c&=&2.7 \times 10^{15}\, {\rm Hz\,} (1+z)^{-1/2} \epsilon_{B,-2}^{-3/2} 
E_{52}^{-1/2}n_1^{-1} t_{day}^{-1/2} (1+x)^{-2}, \\
\label{fmax}
f_{max}&=&2.6\, {\rm mJy\,} (1+z) \epsilon_{B,-2}^{1/2} E_{52} n_1^{1/2}
D_{L,28}^{-2},
\end{eqnarray}
where $D_{L,28}$ is the luminosity distance in units of $10^{28}\,{\rm cm}$.
The coefficients in these equations were taken from the synchrotron
spectrum calculation of \cite{gps99} and are slightly dependent on
$p$. The numerical values here are quoted for $p=2.2$.

Combining Eqs. (\ref{x}), (\ref{nua})--(\ref{fmax}), we can now derive
an interesting constraint on the observable afterglow parameters:
\begin{equation}
C\equiv
0.06 (1+z)^4 t_{day}^4 D_{L,28}^{-2}
\eta
\left( \frac {\nu_a} {\rm GHz} \right)^{\frac {10} {3}} 
\left( \frac {\nu_m} {10^{13}\,{\rm Hz}} \right)^{\frac {13} {6}} 
\left( \frac {\nu_c} {10^{14}\,{\rm Hz}} \right)^{\frac 3 2} 
\left( \frac {F_{\nu_m}} {\rm mJy} \right)^{-1} 
=\frac x {(1+x)^2}.
\label{combination}
\end{equation}
The expression on the right hand side has a maximum value of
$1/4$. Thus, Eq. (\ref{combination}) above gives us a theoretical
constraint on this combination of the observed break frequencies and
peak flux for any afterglow spectrum, completely independent of its
underlying physical parameters!  Note that $\eta$ is also an
observable quantity, since it is given by $\eta =
\min[(\nu_c/\nu_m)^{(2-p)/2},\ 1]$ The only assumption made in deriving
this constraint is that we are seeing the synchrotron emission from a
power-law distribution of electrons accelerated by a relativistic
shock wave.

For fast cooling, we can write a similar combination.  It is simpler,
however, to use same combination $C$ as above, in which we substitute
the ``adjusted'' frequency given by 
\begin{equation}
\nu_a^{slow}=\nu_a^{fast} \left( \frac {\nu_c} {\nu_m} \right) ^{1/2},
\label{nuaslow}
\end{equation}
instead of the observed self absorption frequency $\nu_a^{fast}$.

To illustrate how this constraint can be used, let's consider
GRB~970508, which has probably the best studied afterglow to
date. \cite*{gea98} described the observed broad band spectrum of that burst
and have shown it to be in good agreement with the synchrotron
spectrum of \cite{spn98}. Their fit shows that at $t=12.1\,$days
$\nu_c\cong 1.6 \times 10^{14}\,$Hz $\nu_m \cong 8.6 \times
10^{10}\,$Hz, $\nu_a \cong 3.1\,$GHz and $f_{max}=1.7\,$mJy and
$p\cong 2.2$.
Using the break frequencies of GRB~970508 as given by \cite{gea98}, we
see that Eq. (\ref{combination}) gives $C \cong 4 \gg 1/4$.  Though
this result seems to suggest that the observed spectrum could not be
produced by synchrotron emission, a more likely explanation is that
the values of the break frequencies are not accurate. Both $\nu_m$ and
$\nu_c$ are rather weakly constrained by the observations, and a factor
of a few could easily resolve this disagreement.

We recommend the following procedure to determine the afterglow
parameters from a snapshot of the spectrum. First, one calculates the
combinations of observable as given in Eq. \ref{combination}
(with the correction of Eq. \ref{nuaslow}, if $\nu_c<\nu_a$). If this
combination of observable $C$ is above $1/4$, then there is no
consistent solution. If $C<1/4$ there are two solutions which can be
found in the following way.  First we solve Eq. (\ref{combination}) to
give the two possible values of $x$:
\begin{equation}
x_1=\frac {1-2C-\sqrt{1-4C}} {2C} \cong C \ll 1
\end{equation}
and
\begin{equation}
x_2=\frac {1-2C+\sqrt{1-4C}} {2C} \cong \frac 1 {C} \gg 1.
\end{equation}
The two solutions for the physical parameters describing the afterglow
are then given by:
\begin{eqnarray}
E_{52}= 0.23
\left( \frac {\nu_a} {\rm GHz} \right)^{-\frac {5} {6}} 
\left( \frac {\nu_m} {10^{13}\,{\rm Hz}} \right)^{-\frac {5} {12}} 
\left( \frac {\nu_c} {10^{14}\,{\rm Hz}} \right)^{\frac 1 4} 
\left( \frac {F_{\nu_m}} {\rm mJy} \right)^{\frac 3 2} 
t_{day}^{\frac 1 2} 
(1+z)^{-2}
D_{L,28}^3
(1+x)^{\frac 1 2} \\ 
\ee= 0.23
\left( \frac {\nu_a} {\rm GHz} \right)^{\frac {5} {6}} 
\left( \frac {\nu_m} {10^{13}\,{\rm Hz}} \right)^{\frac {11} {12}} 
\left( \frac {\nu_c} {10^{14}\,{\rm Hz}} \right)^{\frac 1 4} 
\left( \frac {F_{\nu_m}} {\rm mJy} \right)^{-\frac 1 2}  
t_{day}^{-\frac 3 2}
(1+z)^1
D_{L,28}^{-1}
(1+x)^{\frac 1 2} \\ 
\eB= 3.9
\left( \frac {\nu_a} {\rm GHz} \right)^{-\frac {5} {2}} 
\left( \frac {\nu_m} {10^{13}\,{\rm Hz}} \right)^{-\frac {5} {4}} 
\left( \frac {\nu_c} {10^{14}\,{\rm Hz}} \right)^{-\frac 5 4} 
\left( \frac {F_{\nu_m}} {\rm mJy} \right)^{\frac 1 2}  
t_{day}^{\frac 5 2}
(1+z)^{-3}
D_{L,28}^1
(1+x)^{-\frac 5 2} \\ 
n_1= 0.0072
\left( \frac {\nu_a} {\rm GHz} \right)^{\frac {25} {6}} 
\left( \frac {\nu_m} {10^{13}\,{\rm Hz}} \right)^{\frac {25} {12}} 
\left( \frac {\nu_c} {10^{14}\,{\rm Hz}} \right)^{\frac 3 4} 
\left( \frac {F_{\nu_m}} {\rm mJy} \right)^{-\frac 3 2} 
t_{day}^{-\frac 7 2}
(1+z)^5
D_{L,28}^{-3}
(1+x)^{\frac 3 2}
\label{inversion}
\end{eqnarray}

Since $1+x_1 \cong 1$ one of these solutions degenerates to that given
in \cite{wig99} with the corrected coefficients of
\cite{gps99}. However, in the second case we have $x_2 \gg 1$ so that
the resulting values of $E_{52},\ \ee,\ \eB$ and $n_1$ are
significantly different. Substituting $1+x_2 \sim x_2 \sim 1/C$ in
these equations we obtain
\begin{eqnarray}
E_{52}= 0.95 \eta^{-1/2}
\left( \frac {\nu_a} {\rm GHz} \right)^{-\frac {5} {2}} 
\left( \frac {\nu_m} {10^{13}\,{\rm Hz}} \right)^{-\frac {3} {2}} 
\left( \frac {\nu_c} {10^{14}\,{\rm Hz}} \right)^{-\frac 1 2} 
\left( \frac {F_{\nu_m}} {\rm mJy} \right)^{2} 
t_{day}^{\frac 5 2} 
(1+z)^{-4}
D_{L,28}^4, \\ 
\ee= 0.95 \eta^{-1/2}
\left( \frac {\nu_a} {\rm GHz} \right)^{-\frac {5} {6}} 
\left( \frac {\nu_m} {10^{13}\,{\rm Hz}} \right)^{-\frac {1} {6}} 
\left( \frac {\nu_c} {10^{14}\,{\rm Hz}} \right)^{-\frac 1 2} 
t_{day}^{\frac 1 2}  
(1+z)^{-1}, \\ 
\eB= 0.0032 \eta^{5/2}
\left( \frac {\nu_a} {\rm GHz} \right)^{\frac {35} {6}} 
\left( \frac {\nu_m} {10^{13}\,{\rm Hz}} \right)^{\frac {25} {6}} 
\left( \frac {\nu_c} {10^{14}\,{\rm Hz}} \right)^{\frac 5 2} 
\left( \frac {F_{\nu_m}} {\rm mJy} \right)^{-2}  
t_{day}^{-\frac {15} 2}  
(1+z)^7 
D_{L,28}^{-4},\\ 
n_1= 0.51 \eta^{-3/2}
\left( \frac {\nu_a} {\rm GHz} \right)^{\frac {5} {6}} 
\left( \frac {\nu_m} {10^{13}\,{\rm Hz}} \right)^{-\frac {7} {6}} 
\left( \frac {\nu_c} {10^{14}\,{\rm Hz}} \right)^{-\frac 3 2} 
t_{day}^{\frac 5 2} 
(1+z)^{-1}.
\end{eqnarray}
This second solution where IC dominates, was neglected so far. It has
a somewhat higher energy, higher density, higher $\ee$ and
lower $\eB$, than the low-$x$ solution.

How can we determine which of the two solutions is correct at a given
moment? This cannot be done by studying a single snapshot of the
afterglow spectrum. However, the temporal evolution above the cooling
frequency is a little different in the two cases. The decay rate is
slower in the second, IC-dominated, solution. If the decay rate below and
above the cooling frequency is measured to be $t^{-\alpha_1}$ and
$t^{-\alpha_2}$, respectively, then (for constant density of ambient
material) the standard model, which ignores IC emission, predicts
$\alpha_2-\alpha_1=1/4$.  The presence of strong IC cooling will make
the difference smaller, giving $\alpha_2-\alpha_1=1/4-(p-2)/2(4-p)
\cong 0.2$. For a wind-like ambient density profile (see appendix B),
the standard model predicts $\alpha_2-\alpha_1=-1/4$, which changes
to $\alpha_2-\alpha_1=-1/4-(p-2)/(4-p) \cong -0.36$ when the IC
emission dominates the cooling. This difference is not large (except
perhaps in the wind model), and therefore, accurate measurement with
long temporal baselines are needed.

It was recently argued by \cite{frw00} that the total energy contained
in the electrons behind the shock $\ee E$ can be estimated using a
single observation above the cooling frequency. They argued that this
is insensitive to the value of the magnetic field. Their analysis,
which ignores IC, could be explained as follows. Since the electron
energy distribution is flat, i.e. $p\sim 2$, each decade in the
electron distribution contains an amount of energy of the order of the
total energy in all the electrons. Above the cooling frequency all of
this energy is radiated and therefore, the observed $\nu F_{\nu}$
gives an approximate measure of the total electron energy. However, we
have shown that if the ratio $(\ee/\eB)$ is large, most of this energy
is radiated as IC emission rather than synchrotron. Therefore, during
the fast cooling stage the energy estimate suggested by \cite{frw00}
falls short of $\ee E$ by a factor $\sqrt{\ee/\eB}$, while during the
intermediate stage it is short by a factor $\sqrt{\eta \ee/\eB}$,
which is roughly proportional to $\eB^{-0.45}$; and both of these
correction factors depend on the magnetic field.  Thus, \cite{frw00} method
can provide only a lower limit on $\ee E$.

\section{Direct Detection of IC Emission}
\label{direct}
Clearly, the IC emission component can be detected directly only at
frequencies where it dominates over the synchrotron emission. As can
be seen from Fig. \ref{fig1}, for typical parameters this occurs in
the X-ray and $\gamma$-ray bands.  Here we limit our discussion to the
intermediate regime (defined in \S\ref{scIC}) of afterglow evolution,
during which the crossing point between the synchrotron and IC
spectral components, $\nu^{IC}$, generally lies above the synchrotron
cooling frequency $\nu_c$ and between IC break frequencies
$\nu_a^{IC}$ and $\nu_c^{IC}$.  Using the power-law approximation for
both spectral components it is then straight-forward to compute
$\nu^{IC}$, both at early times, when $\nu^{IC} < \nu_m^{IC}$, and at
later times, when $\nu^{IC} > \nu_m^{IC}$.  We obtain the following
expressions:
\begin{equation}
\label{nuic}
\nu^{IC}=\left\{ \begin{array}{ll}
\nu_m^{IC} \left[\frac{\eB}{\ee} \left(\frac{\gc}{\gm}\right)^4 
(2 \gm \gc)^{2-p}\right]^{\frac{3}{2+ 3 p}} \propto 
t^{\frac{3}{2} \frac{(3 p^2 - 8 p - 12)}{(2+3 p) (4-p)}}, & 
\quad {\rm if}\ \nu^{IC} < \nu_m^{IC}; \\
\nu_c^{IC} \frac {\eB} {\ee} (2 \gm \gc)^{2-p}
\propto t^{\frac {3p^2-23p+36} {4(p-4)}}, & 
\quad {\rm if}\ \nu^{IC} > \nu_m^{IC}.
\end{array} \right.
\end{equation}
It is easy to show that the transition between these two regimes occurs when
$\frac{\gc}{\gm} \simeq \left[\frac{\ee}{\eB} \gm^{2
(p-2)}\right]^{\frac{1}{6-p}}$.  Note that it is not necessary to know
a priori which regime is relevant in a given case.  One can simply
compute $\nu^{IC}$ using both expressions given above and then take
the larger value, which will always be the correct one.

To determine whether direct observations of the IC emission is
feasible, it is best to write $\nu^{IC}$ in terms of the underlying
physical parameters describing GRB afterglow.  The general expressions,
though readily computable, are too complicated to give here; instead
we show the results for two characteristic values of $p$. For $p=2.2$ 
(and uniform-density ambient medium), we obtain
\begin{eqnarray}
\label{nuic22}
\nu^{IC}=\max&&\left[2.3 \times 10^{19} \epsilon_{e,.5}^{1.3}
\epsilon_{B,-2}^{0.1} n_1^{-1} t_{day}^{-1.5} (1+z)^{0.5}; \right.\\ \nonumber 
&&\left.\ 1.4 \times 10^{18}  E_{52}^{-1.4} \epsilon_{e,.5}^{-3.7}
\epsilon_{B,-2}^{-0.6} n_1^{-2.3} (1+z)^{-1}\right] {\rm Hz},
\end{eqnarray}
while for $p=2.4$ we get
\begin{eqnarray}
\label{nuic24}
\nu^{IC}=\max&&\left[ 9 \times 10^{18} \epsilon_{e,.5}^{1.2}
\epsilon_{B,-2}^{0.1} n_1^{-1} t_{day}^{-1.4} (1+z)^{0.4}; \right. \\ \nonumber
&& \left.\ 7 \times 10^{16} E_{52}^{-1.6} \epsilon_{e,.5}^{-4.5}
\epsilon_{B,-2}^{-0.75} n_1^{-2.4} t_{day}^{0.3} (1+z)^{-1.3}\right] {\rm Hz}.
\end{eqnarray}
The first and second expressions in parentheses describe early and
late time evolution of $\nu^{IC}$, respectively.  Note that when the
latter expression applies, it may considerably overestimate the value
of $\nu^{IC}$, since it is based on an approximate description of the IC
spectrum (see Fig. \ref{fig1}).    

From Eqs. (\ref{nuic22}) and (\ref{nuic24}) it is clear that
$\nu^{IC}$ reaches its minimum values at a time when $\nu^{IC} =
\nu_m^{IC}$. (Using Eq. (\ref{nuic}) it can easily be shown that this
is a general property of the transition frequency, as long as $p \ga
2.2$.)  Since both {\it Chandra} and {\it XMM} cannot observe hard
X-rays, the IC component can be seen directly only while, say, $\nu^{IC} \la
5\,{\rm keV}$.  This condition places a lower limit on the ambient
density.  For $p=2.2$, we need $n_1 \ga 1.1 E_{52}^{-0.6}
\epsilon_{e,.5}^{-1.6} \epsilon_{B,-2}^{-0.26} (1+z)^{-0.4}$, while
for $p=2.4$ we should have $n_1 \ga 0.4 E_{52}^{-0.6}
\epsilon_{e,.5}^{-1.7} \epsilon_{B,-2}^{-0.27} (1+z)^{-0.5}$, in general 
agreement with \cite{pak00}.

If the ambient density is high enough to satisfy the above conditions
then the following sequence of events should be observed in X-rays
near $5\,{\rm keV}$. The X-ray flux begins to be dominated by IC
emission at time $t \sim 7.5\, {\rm days}\,\epsilon_{e,.5}^{0.9}
\epsilon_{B,-2}^{0.08} n_1^{-0.7} (1+z)^{0.3}$ for $p=2.2$ and $t
\sim 4.3\, {\rm days}\,\epsilon_{e,.5}^{0.9} \epsilon_{B,-2}^{0.07}
n_1^{-0.7} (1+z)^{0.3}$ for $p=2.4$, causing the observed spectral
slope to change from $-p/2$ to $1/3$.  After the transition, the IC
flux increases linearly with time (Eq. \ref{IClightcurve}), producing
a bump in the X-ray lightcurve \citep[see also figure 3 of][]{pak00}.

The observed flux peaks when $\nu_m^{IC}$ moves into the X-ray band at
time $t\sim 7.2\, {\rm days}\,E_{52}^{1/3} \epsilon_{e,.5}^{16/9}
\times \epsilon_{B,-2}^{2/9} n_1^{-1/9} (1+z)^{5/9}$ (independent of $p$).
At the same time the spectral slope changes gradually from $1/3$ to
$-(p-1)/2$, which is still easily distinguishable from the
characteristic $-p/2$ slope of the synchrotron high energy tail. From
this point on, the observed flux evolves as
\begin{equation}
f_{\nu}^{IC} \sim 2 \times 10^{-14} \left(\frac{\nu}{5\,{\rm
keV}}\right)^{-0.6} E_{52}^{1.7} \epsilon_{e,.5}^{2.4} \epsilon_{B,-2}^{0.8}
n_1^{1.1} D_{L,28}^{-2} (1+z)^{1.5} t_{day}^{-1.1}\, {\rm \frac {erg}{s\,cm^{2}\, keV}}, 
\end{equation}
for $p=2.2$ and as
\begin{equation}
f_{\nu}^{IC} \sim 3.2 \times 10^{-14} \left(\frac{\nu}{5\,{\rm
keV}}\right)^{-0.7} E_{52}^{1.8} \epsilon_{e,.5}^{2.8} \epsilon_{B,-2}^{0.9}
n_1^{1.1} D_{L,28}^{-2} (1+z)^{1.6} t_{day}^{-1.3}\, {\rm \frac{erg}{s\,cm^{2}\, keV}},
\end{equation}
for $p=2.4$.
For comparison, ACIS on {\it Chandra} can detect fluxes down to
$4\times 10^{-15}\,{\rm erg\,s^{-1}\,cm^{-2}}$ (in the energy band
$0.4-6\,{\rm keV}$) in a $10^4\,{\rm s}$ observation. Thus, for
reasonable values of the afterglow parameters, the IC emission should
stay visible for several weeks or even months, especially if $E_{52}$
is somewhat larger than unity, as is the case for a reasonable fraction
of the bursts.

\section{Summary}
\label{summary}
In this paper we calculate the detailed spectrum of the inverse
Compton emission from a relativistic shock in the context of GRB
afterglows.  The general shape of this spectral component is very
similar to the primary synchrotron spectrum.  Like the latter, the
spectrum of the IC emission can be roughly approximated by a broken
power-law with break frequencies $\nu_a^{IC}$, $\nu_m^{IC}$, and
$\nu_c^{IC}$ (defined in \S\ref{scspec}).  However, it differs from
the synchrotron spectral component on three major points, summarized
below.  (I) At the low frequency end of the spectrum, $\nu <
\nu_a^{IC}$, the emission increases as $\propto \nu$, rather than as
$\propto \nu^2$ characteristic of the synchrotron spectrum below
$\nu_a$. This may not have major practical consequences, however,
since the low frequency part of the IC spectral component is generally
obscured by the synchrotron high energy tail. (II) The part of the IC
spectrum which lies between $\nu_c^{IC}$ and $\nu_m^{IC}$ extends over
twice as many decades as the corresponding region (between $\nu_m$ and
$\nu_c$) in the synchrotron spectrum. (III) The segments of the IC
spectral component above the peak frequency ($\nu_m^{IC}$ in the slow
cooling regime and above $\nu_c^{IC}$ in the fast cooling regime)
cannot be well described by the pure power-law terms.  The presence of
additional logarithmic terms in this frequency range has the effect of
considerably smoothing out the spectral breaks.  Thus, no sharp
changes in the spectral slope or in the rate of time evolution is
expected during the observation of the IC emission above the peak
frequency.  In contrast, the observed transition between the
synchrotron and Compton component can be sharp and produce quite
dramatic spectral and temporal changes.

In \S\ref{lumratio2} we give a simple prescription for estimating the
importance of IC emission for the total cooling rate of the shocked
gas.  We showed that if the fraction of energy contained in
relativistic electrons, $\epsilon_e$, is larger than that in magnetic
field, $\epsilon_B$, then during the fast cooling stage, IC emission
is greater than the emission due to synchrotron by a factor of
$\sqrt{\epsilon_e/\epsilon_B}$.  This factor decreases very slowly in
the slow cooling regime, since at any given moment electrons located
high enough in the power-law distribution are always cooling. Since
the electron distribution is close to being flat, these electron can
still radiate a significant fraction of the total electron energy.
Thus, the afterglow evolution during the early part of the slow
cooling regime is also dominated by IC emission.  Moreover, we found
that for $\epsilon_e/\epsilon_B > 10$, the IC scattering is likely to
remain the dominant cooling mechanism throughout the entire
relativistic stage.

As long as the IC emission dominates the total cooling rate, it sets
the energy of the cooling break, $\gc$, in the electron energy
distribution, and determines the cooling frequency of the synchrotron
spectral component.  Therefore, the presence of IC emission changes
the values of physical parameters inferred from current afterglow
observations. In \S\ref{infer} we give the revised prescription for
computing the total explosion energy, $E$, the fractions of energy in
electrons and magnetic fields, $\ee$ and $\eB$, and the ambient
particle density, $n$, using the observed properties of the
synchrotron spectral component.  We show that at any instance in time
two possible solutions for $E$, $\ee$, $\eB$, and $n$ are allowed, one
where IC cooling is unimportant and another where it dominates the
total emission from the afterglow.  There is no way to distinguish
between the two solutions based on an instantaneous synchrotron
spectrum alone. They differ only in their time evolution of emission
above $\nu_c$ and in the strength of the IC spectral component.

We also obtain an important constraint on the instantaneous values of
the synchrotron break frequencies, peak flux of the synchrotron
component and the redshift of the afterglow in the slow cooling
regime.  We show that due to the presence of the IC cooling rate, a
combination of these parameters must always be smaller than $1/4$.

Finally, we discuss the possibility of detecting the IC emission
component directly.  We show that for reasonable values of the
physical parameters, this component can be detected by Chandra a few
days after the initial burst, as long as the ambient density is
greater than $\sim 1\,{\rm cm^{-3}}$.  Whether or not the IC component
is detected will be apparent from the change in observed spectral slope,
as well as from the bump in the X-ray light curve.

\acknowledgements
RS gratefully acknowledges support from the Sherman Fairchild foundations.
AE was supported by NASA through
Chandra Postdoctoral Fellowship grant \#PF8-10002 awarded by the
Chandra X-Ray Center, which is operated by the SAO for NASA under
contract NAS8-39073.

\appendix
\section{Details of Compton Scattering}

The Thomson optical depth through the shocked medium, $\tau \sim
\sigma_T n R$, is generally very small, of order $10^{-6}$.
In this case the Compton $y$-parameter, $y = \g_e^2 \tau$, determines
the average fractional energy change of seed photons in each
scattering by an electron with Lorenz factor $\g_e$.  When $y < 1$,
only single scattering needs to be considered in computing inverse
Compton emission.  When $y>1$, multiple scattering can be important.
However, in the context of GRB afterglows, a once-scattered
synchrotron photon with initial energy $h \nu$ generally has energy of
order $\g_e^3 h \nu \ga m_e c^2$ in the rest frame of the second
scattering electron. Then Thomson limit no longer applies and the
energy gain in each successive scattering will be reduced due to
electron recoil and to the necessity of using Klein-Nishina scattering
cross section.  We conclude therefore, that multiple scattering of
synchrotron photons can be ignored.

For single scattering, the inverse Compton volume
emissivity for a power-law distribution of scattering electrons is
given by \citep{ryl79}
\begin{equation}
\label{jc}
j^{IC}_{\nu} = 3 \sigma_T \int_{\gm}^{\infty}{d \g N(\g)
\int_0^1{d x\,g(x) \tilde{f}_{\nu_s}(x)}},
\end{equation}  
where $x \equiv \nu/4 \g^2 \nu_s$, $\tilde{f}_{\nu_s}$ is the incident
specific flux at the shock front, and $g(x) = 1+x+2 x \ln{x}-2 x^2$
takes care of the angular dependence of the scattering cross section
in the limit $\g \gg 1$ \citep{blg70}.  However, the quantities which
are easiest to compare with the observations are the flux in the
inverse Compton component, $f^{IC}_{\nu} = j^{IC}_{\nu} \frac{4}{3}
R^3/(4 \pi D^2)$ and the synchrotron flux $f_{\nu_s} =
\tilde{f}_{\nu_s} 4 \pi R^2/(4 \pi D^2)$, where $R$ is the size of the
shocked region and $D$ is the distance to the observer.  Strictly
speaking, both $j$ and $\tilde{f}$ are measured in the rest frame of
the shocked medium, but they transform in the same way to the
observer's frame, so the transformation factors cancel each other out
(as do redshift effects).  Substituting these quantities into
Eq. (\ref{jc}) gives us
\begin{equation}
\label{fc}
f^{IC}_{\nu} = R \sigma_T \int_{\gm}^{\infty}{d \g N(\g)
\int_0^{x_0}{d x\, f_{\nu_s}(x)}},
\end{equation}  
where we approximated $g(x)= 1$ for $0<x<x_0$ to
simplify the integration.  This simplified expression yields correct
behavior for $x\ll 1$.  The value of the parameter $x_0 = \sqrt{2}/3$
is set by ensuring energy conservation, i.e. setting $\int_0^1{x\,
g(x) d x} = \int_0^{x_0}{x\, d x}$ (we approximate $x_0\sim 0.5$ in the 
main text of the paper).  Note that the
scattered flux distribution for monoenergetic photons computed using
this approximation has a maximum at $x=x_0 \approx 0.47$, very close
to that of the exact distribution, which peaks at $x\approx 0.61$.

\subsection{Slow Cooling}

The distribution of seed photons is described by the synchrotron
spectrum, which consists of four power-law segments \citep{spn98}.
Then the inner integral in Eq. (\ref{fc}) yields (to a leading order
in $\nu$ and zeroth order in $\nu_a/\nu_m$ and $\nu_m/\nu_c$):
\begin{equation}
\label{int1}
I = \left\{ \begin{array}{ll}
I_1 \simeq \frac{5}{2} f_{max} x_0 
\left(\frac{\nu_a}{\nu_m}\right)^{\frac{1}{3}} 
\left(\frac{\nu}{4 \g^2 \nu_a x_0}\right), & \nu < 4 \g^2 \nu_a x_0 \\
I_2 \simeq \frac{3}{2} f_{max} x_0 
\left(\frac{\nu}{4 \g^2 \nu_m x_0}\right)^{\frac{1}{3}}, & 
4 \g^2 \nu_a x_0 < \nu < 4 \g^2 \nu_m x_0 \\
I_3 \simeq \frac{2}{(p+1)} f_{max} x_0 
\left(\frac{\nu}{4 \g^2 \nu_m x_0}\right)^{\frac{1-p}{2}}, & 
4 \g^2 \nu_m x_0 < \nu < 4 \g^2 \nu_c x_0 \\
I_4 \simeq \frac{2}{(p+2)} f_{max} x_0 
\left(\frac{\nu_c}{\nu_m}\right)^{\frac{1-p}{2}}
\left(\frac{\nu}{4 \g^2 \nu_c x_0}\right)^{-\frac{p}{2}}, & 
\nu > 4 \g^2 \nu_c x_0. 
\end{array} \right.
\end{equation}
The quantity $f_{max} = f_{\nu_s} (\nu_m)$ is the flux value at the
peak of the synchrotron spectral component.  Note that the inverse
Compton spectrum for monoenergetic electron scattering has the same
frequency dependence above $4 \g^2 \nu_a x_0$ as the input synchrotron
spectrum above the self absorption frequency, $\nu_a$.  In the range
$\nu < 4 \g^2 \nu_a x_0$, the functional form of the differential
cross section dominates and $I_1$ is linear in $\nu$, rather than
quadratic.

The integration over different electron energies again needs to be divided
into four different regimes:
\begin{equation}
\label{int2}
f_{\nu}^{IC} = R \sigma_T \times \left\{ \begin{array}{ll}

\int_{\gm}^{\infty}{d \g N (\g) I_1}, & \nu < \nu_a^{IC}; \\

\left(\int_{\gm}^{\g_{cr} (\nu_a)}{d \g N (\g) I_2} +  
\int_{\g_{cr} (\nu_a)}^{\infty}{d \g N (\g) I_1}\right), & 
\nu_a^{IC} < \nu < \nu_m^{IC}; \\

\left(\int_{\gm}^{\g_{cr} (\nu_m)}{d \g N (\g) I_3} +  
\int_{\g_{cr} (\nu_m)}^{\g_{cr} (\nu_a)}{d \g N (\g) I_2} + \right. & \\
\quad \left. \int_{\g_{cr} (\nu_a)}^{\infty}{d \g N (\g) I_1}\right), & 
\nu_m^{IC} < \nu < \sqrt{\nu_m^{IC} \nu_c^{IC}}; \\

\left(\int_{\gm}^{\g_{cr} (\nu_c)}{d \g N (\g) I_4} +  
\int_{\g_{cr} (\nu_c)}^{\g_{cr} (\nu_m)}{d \g N (\g) I_3} + \right. & \\
\quad \left. \int_{\g_{cr} (\nu_m)}^{\g_{cr} (\nu_a)}{d \g N (\g) I_2} +
\int_{\g_{cr} (\nu_a)}^{\infty}{d \g N (\g) I_1}\right), & 
\nu > \sqrt{\nu_m^{IC} \nu_c^{IC}},
\end{array} \right.
\end{equation}
where $\g_{cr}(\nu) = \sqrt{\nu/4 \nu x_0}$, and the break frequencies
are defined as $\nu_a^{IC}=4 \gm^2 \nu_a x_0$, $\nu_m^{IC}=4 \gm^2
\nu_m x_0$, $\nu_c^{IC}=4 \gc^2 \nu_c x_0$.

Evaluating the integrals in Eq. (\ref{int2}) and again keeping only the 
dominant terms, we obtain
\begin{eqnarray}
\label{fc2} 
f_{\nu}^{IC} &\simeq& R \sigma_T n f_{max} x_0 \\ \nonumber
&\times&\left\{ \begin{array}{ll}

\frac{5}{2} \frac{(p-1)}{(p+1)} \left(\frac{\nu_a}{\nu_m}\right)^{\frac{1}{3}} 
\left(\frac{\nu}{\nu_a^{IC}}\right), & 
\nu < \nu_a^{IC}; \\

\frac{3}{2} \frac{(p-1)}{(p-1/3)} 
\left(\frac{\nu}{\nu_m^{IC}}\right)^{\frac{1}{3}}, & 
\nu_a^{IC} < \nu < \nu_m^{IC}; \\

\frac{(p-1)}{(p+1)} 
\left(\frac{\nu}{\nu_m^{IC}}\right)^{\frac{1-p}{2}}
\left[\frac{4 (p+1/3)}{(p+1)(p-1/3)} + 
\ln{\left(\frac{\nu}{\nu_m^{IC}}\right)}\right], &
\nu_m^{IC} < \nu < \sqrt{\nu_m^{IC} \nu_c^{IC}}; \\

\frac{(p-1)}{(p+1)} 
\left(\frac{\nu}{\nu_m^{IC}}\right)^{\frac{1-p}{2}}
\left[2 \frac{(2 p+3)}{(p+2)} - \frac{2}{(p+1)(p+2)} + 
\ln{\left(\frac {\nu_c^{IC}}{\nu}\right)}\right], &
\sqrt{\nu_m^{IC} \nu_c^{IC}} < \nu < \nu_c^{IC}; \\

\frac{(p-1)}{(p+1)} 
\left(\frac{\nu}{\nu_m^{IC}}\right)^{\frac{-p}{2}}
\left(\frac{\nu_c}{\nu_m}\right)
\left[2 \frac{(2 p+3)}{(p+2)} + \frac{2}{(p+2)^2} + \frac{(p+1)}{(p+2)}
\ln{\left(\frac{\nu}{\nu_c^{IC}}\right)}\right], &
\nu > \nu_c^{IC}.  
\end{array} \right.
\end{eqnarray}
There is no abrupt spectral slope change at $\nu = \sqrt{\nu_m^{IC}
\nu_c^{IC}}$, so that the region with the slope $(1-p)/2$ extends over
twice as many orders of magnitude as the corresponding region in the
seed synchrotron spectrum.

It is important to point out that for $\nu > \nu_m^{IC}$, the presence
of logarithmic terms ensures that a broken power-law is no longer a
good approximation to the correct spectrum. 

The value of the flux at the peak of the inverse Compton component is
given by
\begin{equation}
f_{\nu}^{IC} (\nu_m^{IC}) \simeq 4 \sigma_T R n f_{max} x_0 
\frac{(p-1)(p+1/3)}{(p-1/3)(p+1)^2},
\end{equation}
assuming that $\nu_a \ll \nu_m \ll \nu_c$.

\subsection{Fast Cooling}

In this regime, the inner integral in Eq. (\ref{fc}) evaluates to:
\begin{equation}
\label{int1fc}
I = \left\{ \begin{array}{ll}
I_1 \simeq \frac{5}{2} f_{max} x_0 \left(\frac{\nu_a}{\nu_c}\right)^{\frac{1}{3}} 
\left(\frac{\nu}{4 \g^2 \nu_a x_0}\right), & \nu < 4 \g^2 \nu_a x_0 \\
I_2 \simeq \frac{3}{2} f_{max} x_0 
\left(\frac{\nu}{4 \g^2 \nu_c x_0}\right)^{\frac{1}{3}}, & 
4 \g^2 \nu_a x_0 < \nu < 4 \g^2 \nu_c x_0 \\
I_3 \simeq \frac{2}{3} f_{max} x_0 
\left(\frac{\nu}{4 \g^2 \nu_c x_0}\right)^{-\frac{1}{2}}, & 
4 \g^2 \nu_c x_0 < \nu < 4 \g^2 \nu_m x_0 \\
I_4 \simeq \frac{2}{(p+2)} f_{max} x_0 
\left(\frac{\nu_c}{\nu_m}\right)^{\frac{1}{2}}
\left(\frac{\nu}{4 \g^2 \nu_m x_0}\right)^{-\frac{p}{2}}, & 
\nu > 4 \g^2 \nu_m x_0. 
\end{array} \right.
\end{equation}
Here $f_{max} = f_{\nu_s} (\nu_c)$.

Performing the appropriate integration over $\g$ we get the spectrum of the 
inverse Compton emission: 
\begin{eqnarray}
\label{fc2fc} 
f_{\nu}^{IC} &\simeq& R \sigma_T n f_{max} x_0 \\ \nonumber
&\times&\left\{ \begin{array}{ll}

\frac{5}{6} \left(\frac{\nu_a}{\nu_c}\right)^{\frac{1}{3}} 
\left(\frac{\nu}{\nu_a^{IC}}\right), & 
\nu < \nu_a^{IC}; \\

\frac{9}{10} \left(\frac{\nu}{\nu_c^{IC}}\right)^{\frac{1}{3}}, & 
\nu_a^{IC} < \nu < \nu_c^{IC}; \\

\frac{1}{3} \left(\frac{\nu}{\nu_c^{IC}}\right)^{-\frac{1}{2}}
\left[\frac{28}{15} - 
\ln{\left(\frac{\nu}{\nu_c^{IC}}\right)}\right], &
\nu_c^{IC} < \nu < \sqrt{\nu_c^{IC} \nu_m^{IC}}; \\

\frac{1}{3} 
\left(\frac{\nu}{\nu_c^{IC}}\right)^{-\frac{1}{2}}
\left[2 \frac{(p+5)}{(p+2)(p-1)} - \frac{2 (p-1)}{3 (p+2)} + 
\ln{\left(\frac{\nu_m^{IC}}{\nu}\right)}\right], &
\sqrt{\nu_c^{IC} \nu_m^{IC}} < \nu < \nu_m^{IC}; \\

\frac{1}{(p+2)} 
\left(\frac{\nu}{\nu_m^{IC}}\right)^{\frac{-p}{2}}
\left(\frac{\nu_c}{\nu_m}\right)
\left[\frac{2}{3} \frac{(p+5)}{(p-1)} - \frac{2}{3}\frac{(p-1)}{(p+2)} + 
\ln{\left(\frac{\nu}{\nu_m^{IC}}\right)}\right], &
\nu > \nu_m^{IC},  
\end{array} \right.
\end{eqnarray}
with the break frequencies defined as $\nu_a^{IC}=4 \gc^2 \nu_a x_0$,
$\nu_c^{IC}=4 \gc^2 \nu_c x_0$, $\nu_m^{IC}=4 \gm^2 \nu_m x_0$.  

In this regime, the flux at the peak of the inverse Compton component is 
\begin{equation}
f_{\nu}^{IC} (4 \gc^2 \nu_c x_0) \simeq \frac{28}{45} \sigma_T 
R n f_{max} x_0,
\end{equation}
assuming that $\nu_a \ll \nu_c \ll \nu_m$.  

\section{The Wind Case: $\rho \propto R^{-2}$}
For the instantaneous spectrum, there is no difference between the
ambient medium with a wind-like or a constant density profile, since
only the density in front of the shock at the time of observation
matters.  However, the two models clearly have different predictions
for the time evolution of the observed emission. 

Eq. (\ref{ICtoSYN}) and its exact (Eq. \ref{exactx}) and approximate
(Eq. \ref{approxx}) solutions, as well as the expression for $\eta$,
are valid for any density profile, since they come from the analysis
of a ``snapshot'' spectrum.  However, since the time dependence of
$\nu_c/\nu_m$ is different for the wind-like and constant density
profiles, the evolution of $x$ is different in the two cases and so
are the transition times between the three stages.  For the wind
density profile, the importance of IC emission decreases somewhat
faster, since $\nu_c/\nu_m \propto t^2$ in the wind case rather
$\propto t$ as in the constant density case. This makes the difference
between the intermediate regime (slow cooling, IC dominates) and the
final regime (slow cooling with negligible IC) more significant.

For the wind profile, the temporal evolution of $x$ is given by:
\begin{equation}
x = \sqrt{\frac{\ee}{\eB}}
\left(\frac{t}{t_0^{IC}}\right)^{-\frac{(p-2)}{(4-p)}}.
\end{equation}
Assuming $\epsilon_e \gg \epsilon_B$, slow cooling now lasts until
\begin{equation}
t_0^{IC}=3.5 (1+z) 
\left( \frac {\ee} {0.5} \right)^{3/2}
\left( \frac {\eB} {0.01} \right)^{1/2}
 A_\star \, {\rm days}
\end{equation}
which is typically longer than in the constant density case. The IC
dominated stage lasts until
\begin{equation}
t^{IC}=2.5\, {\rm years} \left( \frac {\ee/0.5} {\eB/0.01} \right)^
{\frac {4-p} {p-2}} \left( \frac {\ee} {0.5} \right)^{3/2} \left( \frac
{\eB} {0.01} \right)^{1/2} A_\star
\end{equation}
This estimate is correct only during the relativistic stage
and the importance of IC cooling declines fast, once the shock speed falls
significantly below $c$.

When inferring the parameters from a snapshot spectrum, the estimate
of the total energy and the electron and magnetic energy fraction is
not affected by changes in the density profile.  Again one has to
estimate the combination $C$ as given by Eq. (\ref{combination}) and,
provided that $C<1/4$ (otherwise there is no consistent solution),
solve for $x_1$ and $x_2$, and substitute the values into
Eq. (\ref{inversion}). The equation for the ambient density also
holds, but now it gives the density in front of the shock at the time
of observation.  However, since this density is not constant in time,
the resulting number is not very useful.  It is better to 
compute the normalization coefficient for the
$n \propto R^{-2}$ law.  Using the \cite{chl99} notation, 
$\rho=5\times 10^{11}{\rm gr\,cm^{-3}} A_\star R_{\rm cm}^{-2}$, we get
\begin{equation}
A_{\star}=3 \times 10^{-3}
\left( \frac {\nu_a} {\rm GHz} \right)^{\frac {5} {3}}
\left( \frac {\nu_m} {10^{13}\,{\rm Hz}} \right)^{\frac {5} {6}}
\left( \frac {\nu_c} {10^{14}\,{\rm Hz}} \right)^{\frac 1 2}
t_{day}^{2}
(1+z)
(1+x),
\end{equation}
which is similar to the expression of \cite{chl99} with an extra
factor $(1+x)$.  

For the new IC-dominated solution, given by $x_2\simeq 1/C \gg 1$,
we obtained:
\begin{equation}
A_{\star}=0.05 \eta^{-1}
\left( \frac {\nu_a} {\rm GHz} \right)^{-\frac {5} {3}}
\left( \frac {\nu_m} {10^{13}\,{\rm Hz}} \right)^{-\frac {4} {3}}
\left( \frac {\nu_c} {10^{14}\,{\rm Hz}} \right)^{-1}
\left( \frac {F_{\nu_m}} {\rm mJy} \right)
t_{day}^{-2}
(1+z)^{-3}
D_{L,28}^{2}
\end{equation}
As in the constant density case, this new solution, corresponds to
a higher ambient density than the low-$x$ solution.

\bibliography{grb}

\vfill\eject

\includegraphics{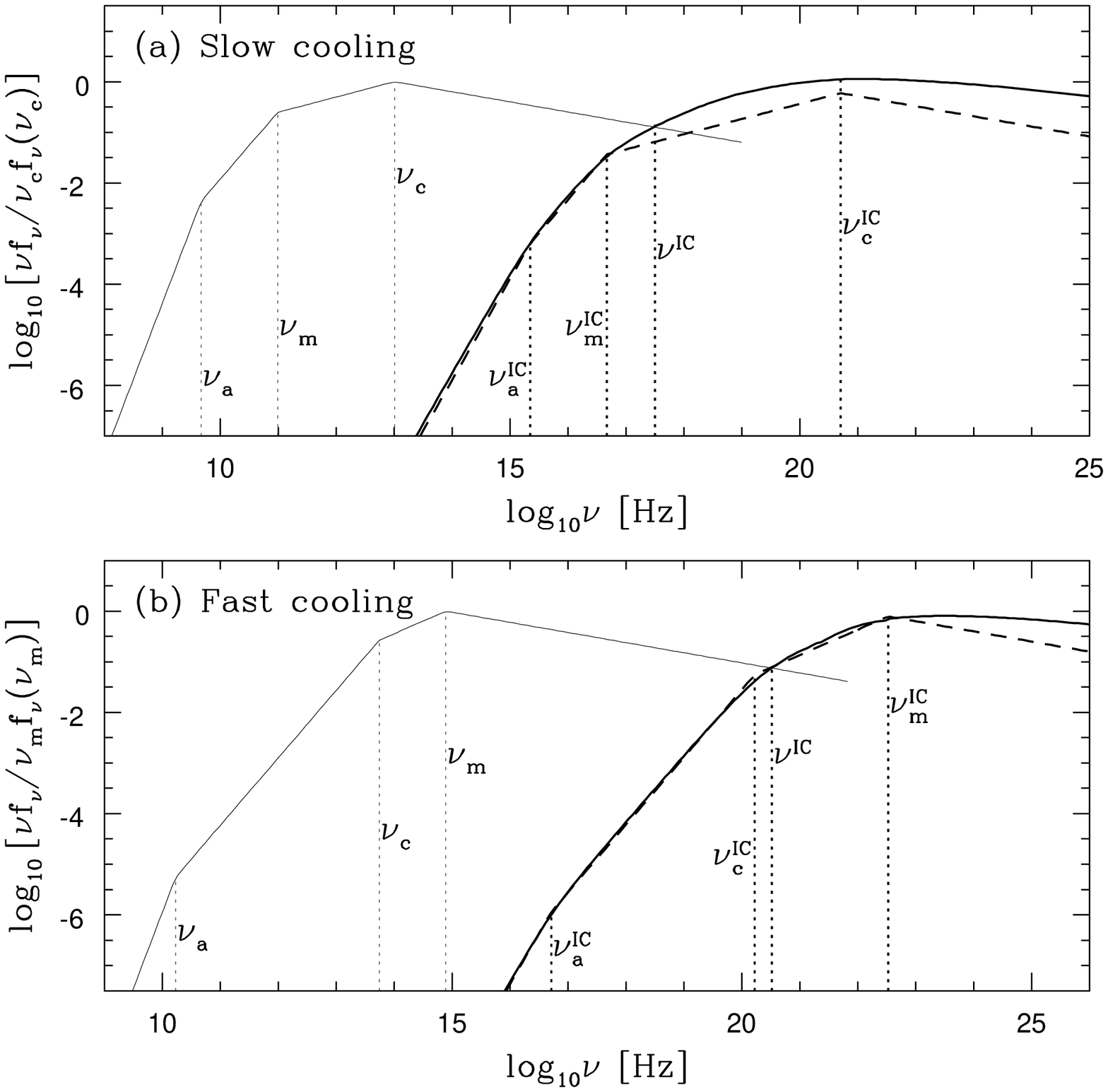}
\figcaption{\label{fig1} Total energy spectrum of a GRB afterglow,
calculated using the following parameters: $p=2.4$, $\ee=0.5$,
$\eB=0.01$, $E_{52}=0.5$, $z=0.5$, $n=3$.  The synchrotron component is shown
as a thin solid line and the inverse Compton component as a heavy
solid line.  A broken power-law approximation to the inverse Compton
spectrum, normalized using Eq. (\ref{fratio}), is plotted as a dashed
line for comparison.  Panel (a) shows the spectrum at $t=12\,{\rm
days}$, when the afterglow is in the slow cooling regime.  Panel (b) shows
the spectrum in the fast cooling regime, computed for $t=43\,{\rm min}$.}
\end{document}